\documentclass[preprint, prd]{revtex4-1} 
\usepackage{graphicx}
\usepackage{lineno} 
\usepackage{dcolumn}
\usepackage{bm}
\usepackage{epsfig}
\usepackage{subfigure}
\usepackage{multirow}
\usepackage{amsmath}
\usepackage{setspace}
\usepackage{longtable}

\newcommand{\ket}[1]{\ensuremath{\left|#1\right\rangle}}

\usepackage{verbatim}

\begin{document}
\title{The sensitivity of the ICAL detector at India-based Neutrino Observatory to neutrino oscillation parameters}
\author{Daljeet Kaur}\email{daljeet.kaur97@gmail.com}
\author{Md. Naimuddin}\email{nayeemsworld@gmail.com}
\author{Sanjeev Kumar}\email{sanjeev3kumar@gmail.com}
\affiliation{Department of Physics and Astrophysics, University of Delhi,\\ 
Delhi-110007, India.}

\date{\today}

\begin{abstract}
The India-based Neutrino Observatory (INO) will host a 50 kt magnetized iron calorimeter (ICAL) detector that will be able to detect muon tracks and hadron showers produced by Charged-Current muon neutrino interactions in the detector. The ICAL experiment will be able to determine the precision of atmospheric neutrino mixing parameters and neutrino mass hierarchy using atmospheric muon neutrinos through earth matter effect. In this paper, we report on the sensitivity for the atmospheric neutrino mixing parameters ($\sin^{2}\theta_{23}$ and $|\Delta m^{2}_{32}|$) for the ICAL detector using the reconstructed neutrino energy and muon direction as observables. We apply realistic resolutions and efficiencies obtained by the ICAL collaboration with a GEANT4-based simulation to reconstruct neutrino energy and muon direction. Our study shows that using neutrino energy and muon direction as observables for a $\chi^{2}$ analysis, ICAL detector can measure $\sin^{2}\theta_{23}$ and $|\Delta m^{2}_{32}|$ with 13\% and 4\% uncertainties at 1$\sigma$ confidence level for 10 years of exposure.
\end{abstract}


\maketitle

\section{Introduction}
 Accumulation of more and stronger evidences of neutrino oscillations from several outstanding neutrino oscillation experiments with atmospheric \cite{sk_osc, sk_atm, minos_atm, T2K_atm, pingu}, solar \cite{hom, Gallex, SAGE, Borexino, SNO1, SNO2, sk1, sk2, sk3, sk4, sk_sol} and reactor \cite{kam, DC, daya, reno}  neutrinos have proven beyond any doubt that neutrinos have mass and they oscillate. In fact, neutrino oscillations were the first unambiguous hint for physics beyond the standard model of elementary particles. In the standard framework of oscillations, the neutrino flavor states are linear superpositions of the mass eigenstates with well defined masses:
\begin{eqnarray}
\ket{\nu_{\alpha}} =\sum_{i}U_{\alpha i} \ket{\nu_{i}},
\label{eq:first}
\end{eqnarray}
where $U$ is the $3 \times 3$ unitary Pontecorvo-Maki-Nakagawa-Sakata (PMNS) \cite{pmns1,pmns2} mixing matrix. In the standard parameterisation \cite{pdg}, PMNS matrix is given as:
\small
\begin{eqnarray}
U^{PMNS}=
\left (
\begin{array}{ccc}

c_{12}c_{13} & s_{12}c_{13} & s_{13}e^{-i \delta}  \\
-s_{12}c_{23}-c_{12}s_{23}s_{13}e^{i \delta}& c_{12}c_{23}-s_{12}s_{23}s_{13}e^{i \delta} & s_{23}c_{13}\\
s_{12}s_{23}-c_{12}c_{23}s_{13}e^{i \delta}& -c_{12}s_{23}-s_{12}c_{23}s_{13}e^{i \delta} & c_{23}c_{13}
\end{array} 
\right).
\label{eq:PMNS}
\end{eqnarray}
\normalsize
Here, $c_{ij}=\cos\theta_{ij}$, $s_{ij}=\sin\theta_{ij}$ and $\delta$ is the Charge-Parity (CP) violating phase.
The neutrino mixing matrix $U^{PMNS}$ can be parameterized in terms of three mixing angles $\theta_{12}$, $\theta_{23}$ and $\theta_{13}$, and a CP phase $\delta$ so that neutrino oscillations are determined in terms of these parameters as well as two mass squared differences, $\Delta m^{2}_{21}$ and $\Delta m^{2}_{32}$. The neutrino oscillations are only sensitive to differences of the squares of three neutrino masses $m_{1}$, $m_{2}$ and $m_{3}$: 
$\Delta m^{2}_{21} = m_{2}^{2}-m_{1}^{2}$ and 
$\Delta m^{2}_{32}=m_{3}^{2}-m_{2}^{2}$.
 The parameters $\Delta m^{2}_{21}$ and $\theta_{12}$ are constrained by the solar neutrino experiments. 
 The atmospheric oscillation parameters $\Delta m^{2}_{32}$ and $\theta_{23}$ were first constrained by Super-Kamiokande \cite{sk_osc} experiment. The sensitivities of these atmospheric parameters were further improved by the MINOS \cite{minos_atm} and  T2K \cite{T2K_atm} experiments. Recently, DAYA Bay \cite{daya}, RENO \cite{reno}, MINOS\cite{minos_th13} and T2K \cite{T2K_th13} experiments have measured the third mixing angle $\theta_{13}$. With the conclusive measurement of relatively large and non-zero value of $\theta_{13}$ from these experiments, the effect of CP violation in neutrino oscillations is expected to be within the reach of future neutrino experiments. This discovery has also opened up a possibility to answer the various unsolved issues of current neutrino physics like whether the neutrino mass hierarchy is normal ($m^{2}_{3} >  m^{2}_{2}$) or inverted  ($m^{2}_{3} < m^{2}_{2}$), what the octant of $\theta_{23}$ is (whether $\theta_{23}$ $<$ $45^{\circ}$ or $\theta_{23}$ $>$ $45^{\circ}$), what is the value of CP violating phase $\delta$, etc. Apart from these questions, the higher precision measurement of current neutrino mixing angles and mass square differences is also very important. The current best fit values and errors in the oscillation parameters on the basis of global neutrino analyses \cite{global1, global2, global3, global4} are summarised in Table \ref{osc_tb}.
A large number of neutrino experiments are ongoing and proposed to achieve the above mentioned goals \textit{viz.} MINOS \cite{minos_tdr}, T2K \cite{t2k_tdr}, INO \cite{ino_tdr, ino_web}, PINGU \cite{pingu}, Hyper-Kamiokande \cite{hk}, NO$\nu$A \cite{nova} etc. Present work is  focused only on the magnetised Iron CALorimeter (ICAL) detector at the India-based Neutrino Observatory (INO). 
\begin{table}
\begin{center}
\begin {tabular}{|c|c|c|}
\hline
Oscillation parameters & True values & Marginalisation range  \\
\hline 
$ \sin^2(2\theta_{12})$ &  0.86 & Fixed \\
$ \sin^2(\theta_{23})$ &  0.5 &  0.4-0.6 ($3\sigma$ range)\\
$ \sin^2(\theta_{13})$ &  0.03 &  0.02-0.04 ($3\sigma$ range )\\
$\Delta m^{2}_{21}$ (eV$^{2}$) &  7.6 $\times$ $10^{-5}$& Fixed\\
 $\Delta m^{2}_{32}$ (eV$^{2}$) & 2.4 $\times$ $10^{-3}$ & (2.1-2.6) $\times$ $10^{-3}$ $(3\sigma~range )$\\
 $\delta$ & 0.0 & Fixed \\
 \hline
\end {tabular}
\caption{\label{osc_tb} Current best fit values of oscillation parameters and their 3 standard deviation range.} 
\end{center}
\end{table}
 
India-based Neutrino Observatory (INO) is a proposed underground laboratory located at Theni district in southern India. The ICAL detector at INO will study mainly the atmospheric muon neutrinos and anti-neutrinos.
Because of being magnetised, the ICAL detector can easily distinguish between atmospheric $\nu_{\mu}$ and $\bar\nu_{\mu}$ by identifying the charge of muons produced in Charged-Current interactions of these neutrinos in the detector.
Re-confirmation of atmospheric neutrino oscillations, precision measurement of oscillation parameters and the determination of neutrino mass hierarchy through the observation of earth matter effects in atmospheric neutrinos are the primary physics goals of the INO-ICAL experiment. Matter effects in neutrino oscillations are sensitive to the sign of $\Delta m^{2}_{32}$. Though the ICAL experiment is insensitive to $\delta$ \cite{anush}, it has been observed that INO mass hierarchy results together with other experiments can help to determine the value of $\delta$ \cite{surbavati}.

In this paper, we present the precision measurement of atmospheric neutrino oscillation parameters ($|\Delta m^{2}_{32}|$ and $\sin^{2} \theta_{23}$) in 3-flavor mixing scheme through the earth matter effect for ICAL detector at INO. 

The precision study of these parameters is important to assess the ICAL capability vis-a-vis other experiments. The sensitivity of the ICAL experiment for these oscillation parameters has already been studied by binning the events in the muon energy and muon angle, using realistic muon resolutions and efficiencies \cite{trk}. Here, however, we use a different approach to determine the sensitivity of ICAL detector for the atmospheric neutrino mixing parameters.  When atmospheric $\nu_{\mu}(\bar\nu_{\mu})$ interact with the ICAL detector, it produces $\mu^{+}(\mu^{-})$ and shower of hadrons. In order to extract the full information about the parent neutrino, information of muons along with that of hadrons is used in the analysis. Recently, it has been shown that including  hadron information together with the muon events improves the ICAL potential for the measurement of neutrino mass hierarchy \cite{hdphy,3dpaper}. Since the neutrino energy cannot be measured directly, therefore, in the analysis presented here, the neutrino energy is obtained by adding the energy deposited by the muons and hadron inside the ICAL detector. We then use this neutrino energy ($E_{\nu}$) and muon angle ($\cos\theta_{\mu}$) as observables for the $\chi^{2}$ estimation. An earlier analysis have used hadron information, but with constant resolutions to obtain the neutrino energy \cite{samanta-smirnov}. In the present work, we show the ICAL potential for the neutrino oscillation parameters using effective realistic ICAL detector resolutions.  

The analysis starts with the generation of the neutrino events with NUANCE \cite{nuance} and then events are binned into $E_{\nu}$ and $\cos\theta_{\mu}$ bins. Various resolutions and efficiencies obtained by INO collaboration from a GEANT4 \cite{geant} based simulation are applied to these binned events in order to reconstruct the neutrino energy and muon direction. Finally, a marginalised $\chi^{2}$ is estimated over the allowed ranges of neutrino parameters, other than $\theta_{23}$ and $|\Delta m^2_{32}|$, after including the systematic errors.

 \section{ The ICAL detector and atmospheric neutrinos}
\label{ical}
The ICAL detector at INO \cite{ino_tdr,ino_web} will be placed under approximately 1 km of rock cover from all directions to reduce the cosmic background.  The detector will have three modules, each of size 16 m $\times$ 16 m $\times$ 14.45 m in x, y and z directions respectively. ICAL consists of 151 horizontal layers of 5.6 cm iron plates with 4 cm of gap between two successive iron layers. Gaseous detectors called Resistive Plate Chambers (RPCs) of dimension 2 m $\times$ 2 m will be used as active detector element and will be interleaved in the iron layer gap. RPC detectors are known for their good time resolution ($\sim$ 1 ns) and spatial resolution ($\sim$ 3 cm). The RPCs provide two dimensional readouts through the external copper pick up strips placed above and below the detector. Total mass of the detector is approximately 50 kt which would provide the statistically significant data to study the weakly interacting neutrinos. A magnetic field of upto 1.5 tesla will be generated through the solenoidal coils placed around the detector. The ICAL detector can easily identify the charge of muons due to this applied magnetic field, and hence, can easily distinguish between neutrinos and anti-neutrinos.

Atmospheric muon neutrinos and anti-neutrinos are the main sources of events for the ICAL detector. When cosmic rays interact with the earth's upper atmosphere, they produce pions which further decay into leptons and corresponding neutrinos. The dominant channels of the decay chain producing atmospheric neutrinos, are
\begin{eqnarray}
\label{atm_int}
\pi^{+} \rightarrow \mu^{+} \nu_{\mu},       ~~~~~~~ \mu^{+} \rightarrow e^{+}\nu_{e}\bar{\nu_{\mu}}, \nonumber\\
\pi^{-} \rightarrow \mu^{-} \bar{\nu_{\mu}}, ~~~~~~~ \mu^{-} \rightarrow e^{-}\bar{\nu_{e}}\nu_{\mu}.
\end{eqnarray}
 Atmospheric neutrinos come in both $\nu_{\mu}$ and $\nu_{e}$ ( $\bar\nu_{\mu}$ and $\bar\nu_{e}$) flavors with the $\nu_{\mu}$ flux almost double that of $\nu_{e}$ flux. 
Due to the large flight path and the wide coverage of the energy range (from few hundred MeV to TeV), atmospheric neutrinos play an important role in studying the neutrino oscillations. Neutrinos and anti-neutrinos interact differently with earth matter. We can use this special feature to measure the sign of $\Delta m^{2}_{32}$, and hence, the correct mass ordering.

\section{Analysis}
\label{process}

The atmospheric neutrino events are generated with the available 3-dimensional neutrino flux provided by HONDA et al.\cite{honda} using ICAL detector specifications. The interactions of atmospheric muon neutrino and anti-neutrino fluxes with the detector target are simulated by the NUANCE neutrino generator for 1000 years of exposure of 50 kt ICAL detector. For the purpose of quoting the final sensitivity we normalise the 1000 years data to 10 years of exposure to keep Monte Carlo fluctuations under control; following the similar approach used in the earlier ICAL analyses \cite{anush, trk}. Only the events generated through Charged-Current (CC) interactions are considered for the present analysis.
 
The neutrino oscillation can be incorporated into the NUANCE code to generate the oscillated neutrino flux at the detector for different values of oscillation parameters. However, this process requires large computational time and resources. Therefore, we simulate the interactions of atmospheric neutrinos with the detector in the absence of oscillations and the effect of oscillations is included by using the re-weighting algorithm described in Refs. \cite{anush,trk}.
For each neutrino event of a given energy $E_{\nu}$ and zenith direction  $\theta_{z}$,  oscillation probabilities are estimated taking earth matter effects into account. 
The path length traversed by neutrinos from the production point to the detector, which is needed as an input parameter in the oscillation probability estimation, is obtained as: 
\begin{equation}
\label{eq:path}
  L = \sqrt{(R_{earth}+R_{atm})^{2}-(R_{earth}\sin\theta_{z})^{2}} - R_{earth}\cos\theta_{z},
 \end{equation} 
where $ R_{earth}$ is the radius of earth and $R_{atm}$ is the average height of the production point of neutrinos in the atmosphere. We have used $R_{earth}$ $\approx$ 6371 km and $R_{atm}$ $\approx$ 15 km. Here, we assume that $\cos\theta_{z}=1$ is the downward and $\cos\theta_{z}=-1$ is the upward direction for incoming neutrinos. The oscillation parameters used in the analysis are listed in Table \ref{osc_tb}.
For the precision measurement studies, we assume normal hierarchy of neutrinos.

In order to separate the muon neutrino and anti-neutrino events on the basis of their oscillation probabilities, each NUANCE generated unoscillated neutrino event was subjected to the oscillation randomly by applying the event re-weighting algorithm.
Since $\nu_{e}$ may also change flavor to $\nu_{\mu}$ due to oscillations, therefore, to include this contribution, we simulate the interactions of $\nu_{e}$ flux with the ICAL detector in the absence of oscillations using NUANCE and applying the re-weighting algorithm for $\nu_{e} \rightarrow \nu_{\mu}$ channel. Hence, total event spectrum consists of $\nu_{\mu}$ events coming from both the oscillation channels (i.e. $\nu_{\mu} \rightarrow \nu_{\mu}$ and  $\nu_{e} \rightarrow \nu_{\mu}$ ). 

\subsection {ICAL detector resolutions and the neutrino energy reconstruction}
\begin{figure}[ht]
 \centering
 \subfigure[]{
  \includegraphics[width=0.45\textwidth]{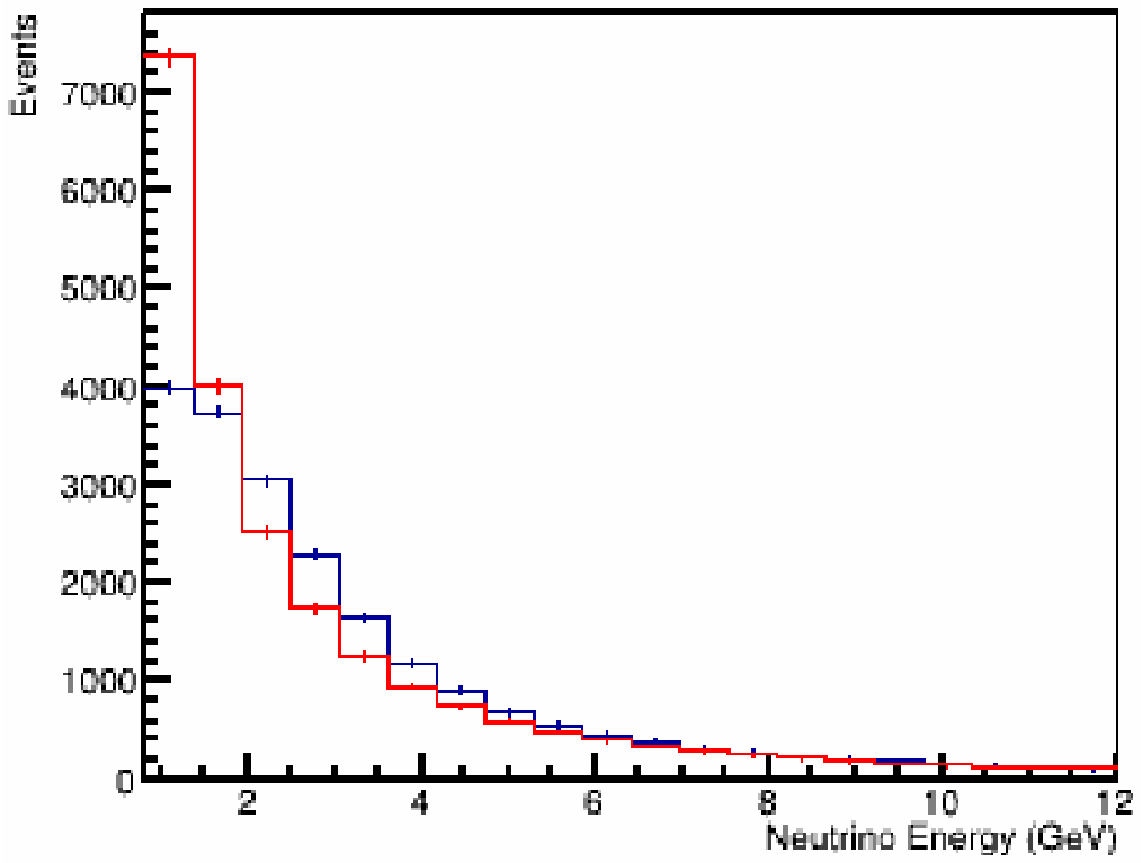}
   }
 \subfigure[]{
  \includegraphics[width=0.45\textwidth]{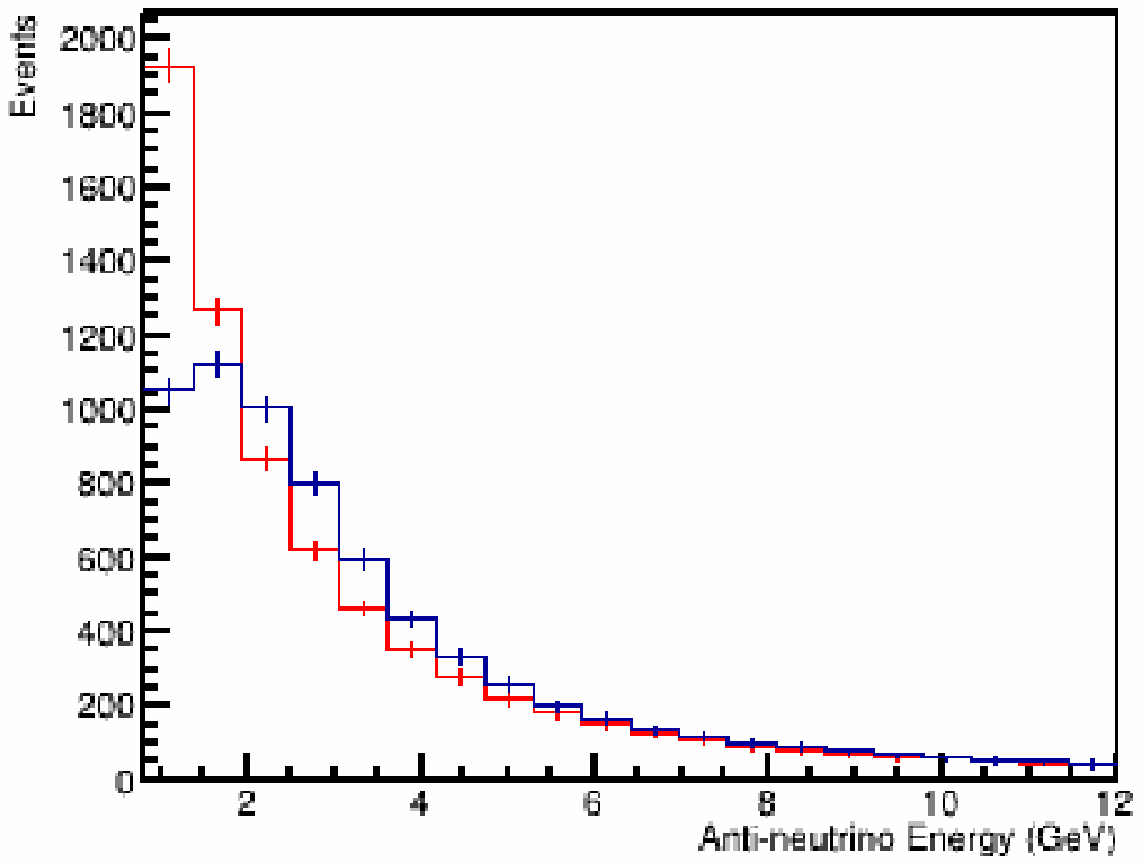}
   }
\caption{\label{recoE1} True neutrino energy (red) and the reconstructed neutrino energy (blue) from nuance simulated data for (a) neutrino events and (b) anti-neutrino events, from $\nu_{\mu}\rightarrow \nu_{\mu}$ oscillation channel}

\end{figure}

\begin{figure}[ht]
 \centering
 \subfigure[]{
  \includegraphics[width=0.45\textwidth]{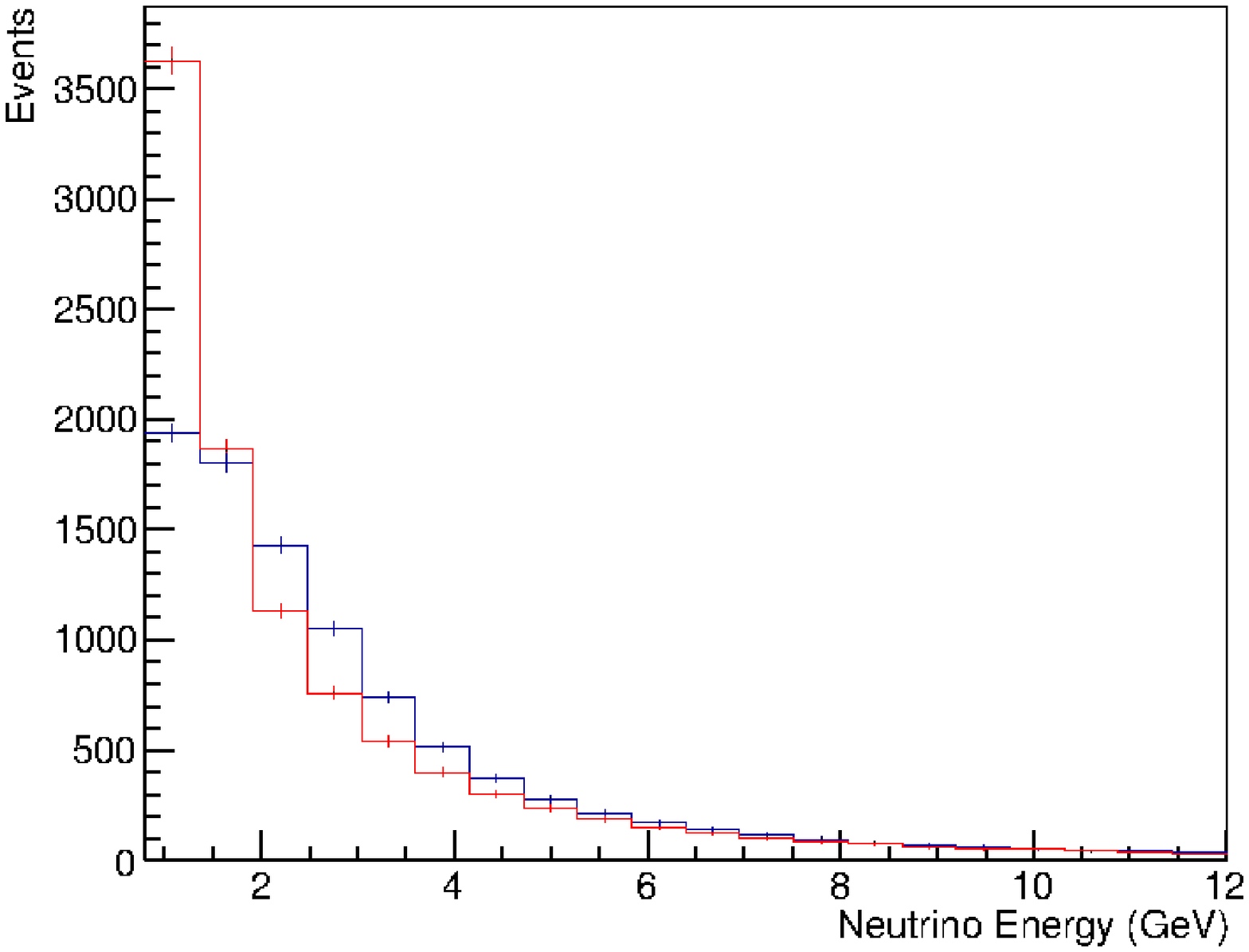}
   }
 \subfigure[]{
  \includegraphics[width=0.45\textwidth]{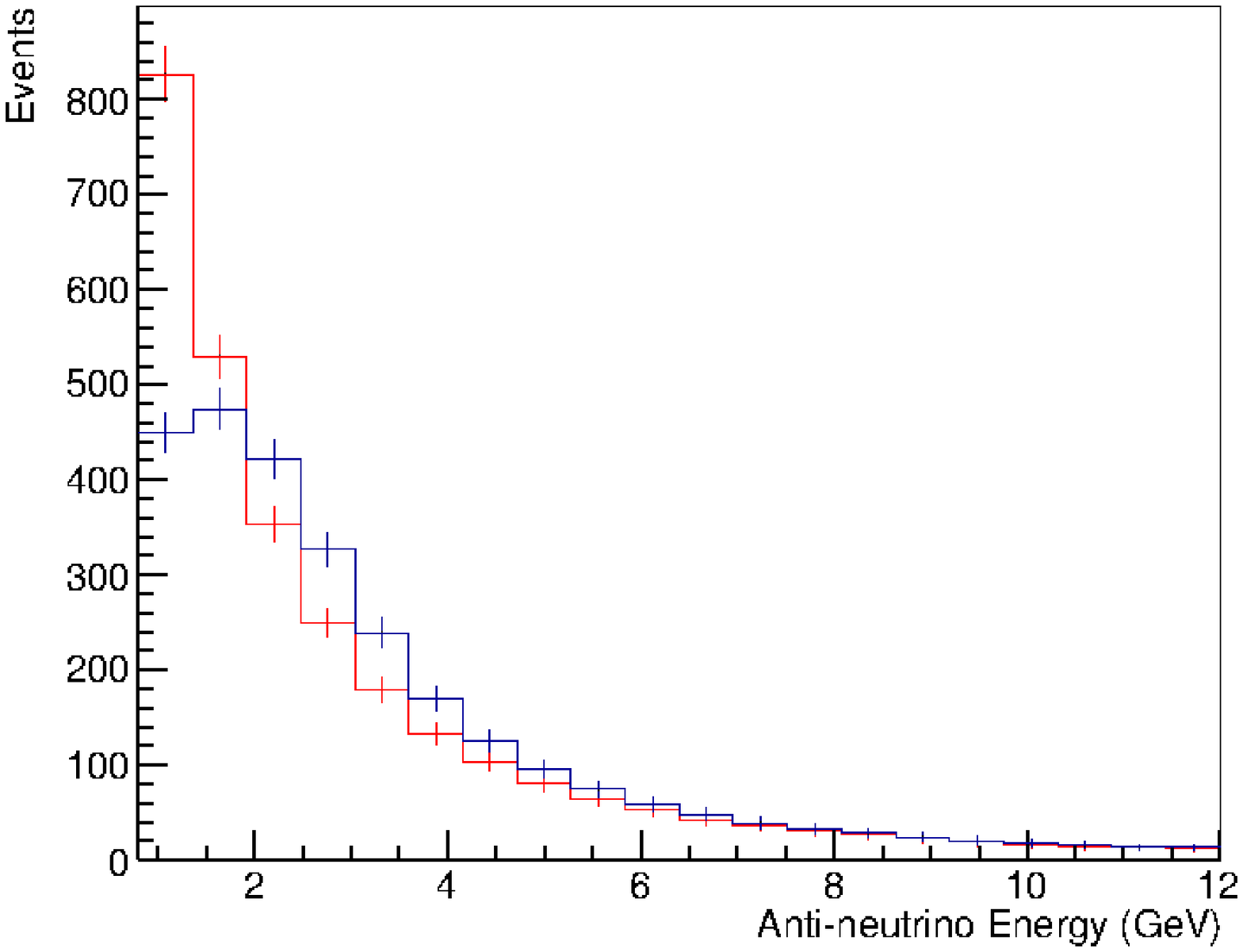}
   }
\caption{\label{recoE2} True neutrino energy (red) and the reconstructed neutrino energy (blue) from nuance simulated data for (a) neutrino events and (b) anti-neutrino events, from $\nu_{e}\rightarrow \nu_{\mu}$ oscillation channel}
\end{figure}

Due to the Charged-Current interaction of the neutrinos in the detector, muons along with the showers of hadrons are produced. Reconstruction of the neutrino energy requires the reconstruction of muon as well as hadron energy. Once we have the reconstructed muon and hadron energies, we directly add them together to get the final reconstructed neutrino energy. Muon and hadron energy resolutions have been obtained by the INO collaboration as function of true energies and true directions of muons or hadrons using a GEANT4 \cite{geant} based code \cite{trk, mureso}. Muons give a clear track of hits inside the magnetised detector, therefore the energy of muons can be reconstructed easily using a track fitting algorithm. It was observed that the muons energy reconstructed by ICAL detector follows Gaussian distribution for $E_{\mu} \ge 1$ GeV whereas it follows Landau distribution for  $E_{\mu} < 1$ GeV. On the other hand, hadrons deposit their energies in a shower like pattern. Total energy deposited by the hadron shower ($E^{\prime}_{had} = E_{\nu}-E_{\mu}$) has been used to calibrate the detector response. It has been found that hadron hit patterns follow Vavilov distribution \cite{vavfunc}. The hadron energy resolution has been fitted as function of $E^{\prime}_{had}$ \cite{hreso}. In the present analysis, muon energy and angular resolutions are implemented by smearing true muon energy and direction of each $\mu^{+}$ and $\mu^{-}$ event using the ICAL muon resolution functions \cite{mureso}. True hadron Energies are smeared using ICAL hadron resolution functions \cite{hreso}. Reconstructed neutrino energy is then taken as the sum of smeared muon and hadron energy. 
Fig~.\ref{recoE1} shows the true and reconstructed neutrino and anti-neutrino energies obtained from the $\nu_{\mu}\rightarrow \nu_{\mu}$ channel while Fig.~\ref{recoE2} shows the same for $\nu_{e}\rightarrow \nu_{\mu}$ channel. It can be seen that at lower incoming neutrino energies ($E_{\nu}\le1$), reconstruction of neutrino energy at ICAL is poor due to the effect of detector resolutions in this range.

 Since the muon direction reconstruction is extremely good for ICAL, and hadron direction information not available yet, we have used  the reconstructed muon directions in the final analysis.



The reconstruction and charge identification efficiencies (CID) for $\mu^{-}$ and $\mu^{+}$ for ICAL detector are included into analysis 
by simply weighted each event with its reconstruction and relative charge identification efficiency. Though the CID efficiencies of ICAL detector are $\ge 90\%$ beyond $E_{\nu}\sim1$ GeV, it is still possible that some muon events (say $\mu^{+}$) are wrongly identified as of the opposite charge particle (say $\mu^{-}$). So, the total number of events reconstructed as $\mu^{-}$ will increase by
\begin{equation}
\label{eq:effeq}
N^{\mu^{-}}= N^{\mu^{-}}_{RC}+(N^{\mu^{+}}_{R}- N^{\mu^{+}}_{RC}),
\end{equation}
where $N^{\mu^{-}}$ are the number of total reconstructed $\mu^{-}$ events.
$N^{\mu^{-}}_{RC}$ are the number of $\mu^{-}$ events reconstructed and correctly identified in charge and $N^{\mu^{+}}_{RC}$ are the number of $\mu^{+}$ events with their respective reconstruction and CID efficiencies folded in; whereas $N^{\mu^{+}}_{R}$ are the number of $\mu^{+}$ events with the reconstruction efficiency only.
Hence, $N_{R}-N_{RC}$ gives the fraction of reconstructed events that have their charge wrongly identified. Total reconstructed $\mu^{+}$ events can be obtained using similar expression with charge reversal.

\section{$\chi^{2}$- Estimation}
\label{chianls}
The sensitivity of the atmospheric neutrino oscillation parameters for ICAL is estimated by minimising the $\chi^{2}$ for the neutrino data simulated for the ICAL detector.
The re-weighted events, with detector resolutions and efficiencies folded in, are binned into reconstructed neutrino energy and muon direction for the determination of $\chi^{2}$. The data has been divided into total 20 varied neutrino energy bins in the range of 0.8 - 10.8 GeV. Since most of the atmospheric neutrino events come below the neutrino energy $E_{\nu} \sim $ 5 GeV, we have a finer energy binning with a bin width of 0.33 GeV from 0.8 to 5.8 GeV with a total of 15 energy bins. The high energy events, i.e. from 5.8 GeV to 10.8 GeV, are divided into total 5 equal energy bins with bin width of 1 GeV. A total of 20 $\cos\theta_{\mu}$ direction bins in the range [-1, 1] with equal bin width, have been chosen. The bin size for the analysis has been optimised such that each bin contains at least one event. The above mentioned binning scheme is applied for both $\nu_{\mu}$ and $\bar\nu_{\mu}$ events. 

We use the maximal mixing, that is, $\sin^{2}\theta_{23}=0.5$ as the reference value. The atmospheric mass square splitting is related to the other oscillation parameters, so for the precision study we have used $\Delta m^{2}_{eff}$, which can be written as \cite{anush, meff}, 
\begin{equation}
\Delta m^{2}_{eff} = \Delta m^{2}_{32}-(\cos^{2}\theta_{12}-\cos \delta\sin\theta_{23}\sin2\theta_{12}\tan\theta_{23})\Delta m^{2}_{21}.
\end{equation}
 The other oscillation parameters ($\theta_{12}$, $\Delta m^{2}_{21}$ and $\delta$) are kept fixed both for observed and predicted events as the marginalisation over these parameters has negligible effects on the analysis results. Since in our analysis, the event samples are distributed in terms of reconstructed neutrino energy and the muon zenith angle bins, we call these events as neutrino-like events that is, we refer to $N^{\mu^-}$ as $N(\nu_\mu)$ and $N^{\mu^+}$ as $N(\bar{\nu}_\mu)$.

The various systematic effects on the $\chi^{2}$ have been implemented through five systematic uncertainties, \textit{viz.} 20 \% error on atmospheric neutrino flux normalisation, 10\% error on neutrino cross-section, a 5\% uncertainty due to zenith angle dependence of the fluxes, an energy dependent tilt error, and an overall 5\% statistical error, as applied in earlier ICAL analyses \cite{anush, trk}. The systematic uncertainties are applied using the method of ``pulls'' as outlined in Ref.\cite{maltoni}. Briefly, in the method of pulls, systematic uncertainties and the theoretical errors are parameterised in terms of set of variables $\zeta_{k}$, called pulls. Due to the fine binning, some bins may have very small number of entries, therefore, we have used the poissonian definition of $\chi^{2}$ given as
 
\begin{equation}
\label{eq:chieq}
 \chi^2(\nu_{\mu}) = min\sum_{i,j}\left(2 (N^{th^{\prime}}_{ij}(\nu_{\mu})-N^{ex}_{i,j}(\nu_{\mu}))+2N^{ex}_{i,j}(\nu_{\mu})
(\ln\frac{N^{ex}_{i,j}(\nu_{\mu})}{N^{th^{\prime}}_{i,j}(\nu_{\mu})})\right)+ \sum_{k}\zeta^{2}_{k}, 
 \end{equation}
 where 
\begin{equation}
\label{eq:evteq}
  N^{th^{\prime}}_{ij}(\nu_{\mu}) = N^{th}_{i,j}(\nu_{\mu})\left(1 + \sum_{k}\pi^{k}_{ij}\zeta_{k}\right). 
 \end{equation}

 Here, $N^{ex}_{ij}$ are the observed number of reconstructed $\mu^{-}$ events, as calculated from Eq. (\ref{eq:effeq}), generated using true values of the oscillation parameters as listed in Table \ref{osc_tb} in $i^{th}$ neutrino energy bin and $j^{th} \cos\theta_{\mu}$ bin.  In Eq. (\ref{eq:evteq}), $N^{th}_{ij}$ are the number of theoretically predicted events generated by varying oscillation parameters, $N^{th^{\prime}}_{ij}$ shows modified events spectrum due to different systematic uncertainties, $\pi^{k}_{ij}$ is the systematic shift in the events of $i^{th}$ neutrino energy bin and $j^{th} \cos\theta_{\mu}$ bin due to $k^{th}$ systematic error. $\zeta_{k}$ is the univariate pull variable corresponding to the $\pi^{k}_{ij}$ uncertainty.
An expression similar to Eq. (\ref{eq:chieq}) can be obtained for $\chi^{2}(\bar\nu_{\mu})$ using reconstructed $\mu^{+}$ event samples.  
 We have calculated  $\chi^2(\nu_{\mu})$ and $\chi^2(\bar\nu_{\mu})$ separately and then these two are added to get total $\chi^2_{total}$ as
 \begin{equation}
\label{eq:chiino}
  \chi^2_{total} =\chi^{2}(\nu_{\mu}) + \chi^{2}(\bar\nu_{\mu}).
 \end{equation}

We impose the recent $\theta_{13}$ measurement as a prior while marginalising over $\sin^{2}\theta_{13}$ as
  \begin{equation}
\label{eq:prioreq}
  \chi^2_{\textit{ical}} =\chi^{2}_{total} +\left(\frac {\sin^{2}\theta_{13}(true)-\sin^{2}\theta_{13}}{\sigma_{\sin^{2}\theta_{13}}}\right)^2. 
 \end{equation}
 The value of $\sigma_{\sin^{2}\theta_{13}}$ was taken as $10\%$ of the true value of $\sin^{2}\theta_{13}$.

Finally, in order to obtain the experimental sensitivity for $\theta_{23}$ and $|\Delta m^{2}_{32}|$, we minimise the $\chi^2_{\textit{ical}}$ function by varying oscillation parameters within their allowed ranges over all systematic uncertainties.

\section {Results}
\begin{figure}[htb]
\center{\includegraphics[height=10cm, width=12cm]{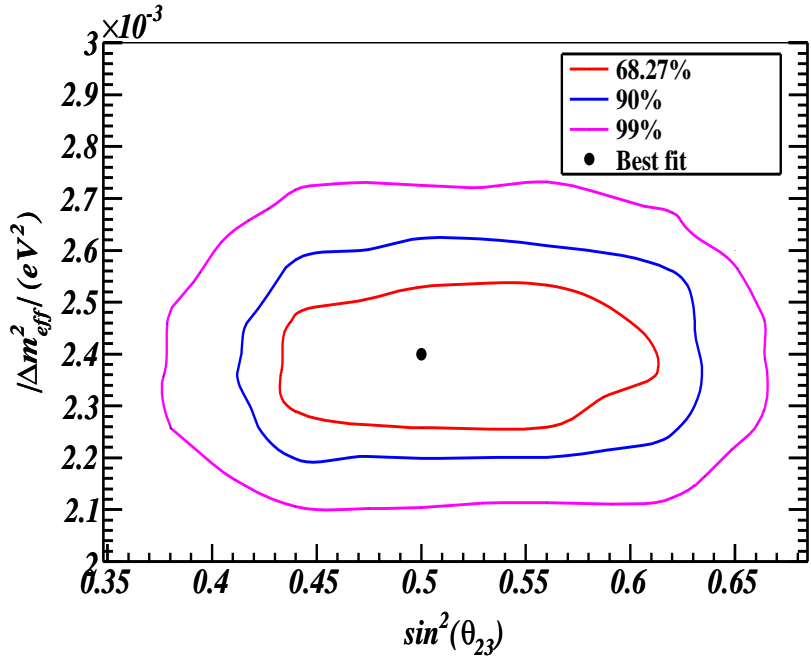}}
\caption{\label{mu_had} Expected sensitivity for $\sin^2(\theta_{23})$ and $|\Delta m^{2}_{eff}|$ at 68$\%$, 90$\%$ and 99$\%$  confidence level  for 10 years exposure of ICAL detector.}
\end{figure}

\begin{figure}[ht]
 \centering
 \subfigure[]{
  \includegraphics[width=0.49\textwidth,height=6cm]{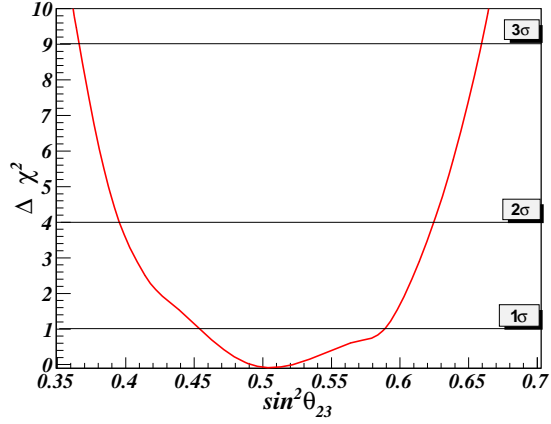}
   }
 \subfigure[]{
  \includegraphics[width=0.49\textwidth,height=6cm]{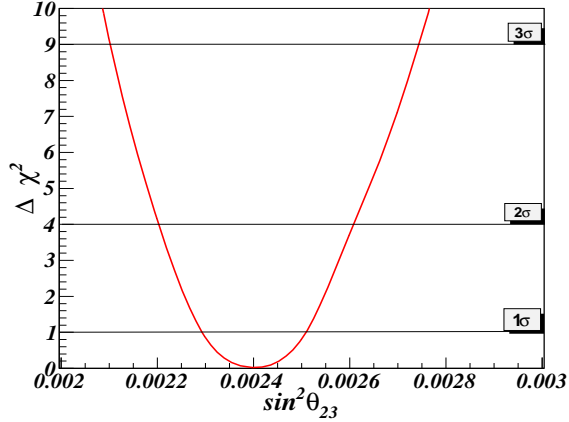}
   }
\caption{\label{sense} (a) $\Delta\chi^{2}$ as a function of test values of $\sin^{2}\theta_{23}=0.5$ and (b) $\Delta\chi^{2}$ as a function of input values of $|\Delta m^{2}_{32}|=2.4 \times 10^{-3}$ eV$^{2}$.}
\end{figure}

The two dimensional confidence region of the oscillation parameters ($|\Delta m^{2}_{eff}|$, $\sin^{2}\theta_{23}$) are determined from $\Delta \chi^{2}_{\textit{ical}}$ around the best fit. The resultant region is shown in Fig. \ref{mu_had}. These contour plots have been obtained assuming $\Delta \chi^{2}_{\textit{ical}} = \chi^{2}_{min} + m$, where $\chi^{2}_{min}$ is the minimum value of $\chi^{2}_{\textit{ical}}$ for each set of oscillation parameters and values of $m$ are taken as 2.30, 4.61 and  9.21 corresponding to 68$\%$, 90$\%$ and 99$\%$ confidence levels \cite{pdg} respectively for two degrees of freedom. Fig. \ref{sense}(a) depicts the one dimensional plot for the measurement of test parameter $\sin^{2}\theta_{23}$ at constant value of  $|\Delta m^{2}_{eff}|=2.4 \times 10^{-3}$ (eV$^{2}$) and Fig. \ref{sense}(b) for the $|\Delta m^{2}_{eff}|$ at constant $\sin^{2}\theta_{23} = 0.5$ at 1$\sigma$, 2$\sigma$ and 3$\sigma$ levels for one parameter estimation \cite{pdg}.

 The precision on the oscillation parameters can be defined as:
 \begin{equation}
 Precision = \frac{P_{max}-P_{min}}{P_{max}+P_{min}},
 \label{pre}
 \end{equation}
where $P_{max}$ and $P_{min}$ are the maximum and minimum values of the concerned oscillation parameters at a given confidence level. The current study shows that ICAL is capable of measuring the atmospheric mixing angle $\sin^{2}\theta_{23}$ with a precision of  13$\%$, 21$\%$ and 27$\%$, at 1$\sigma$, 2$\sigma$ and 3$\sigma$ confidence levels respectively. The atmospheric mass square splitting $|\Delta m^{2}_{32}|$ can be measured with a precision of 4$\%$, 8$\%$ and 12$\%$ at 1 $\sigma$, 2 $\sigma$ and 3 $\sigma$ confidence levels respectively. 
 These numbers show an improvement of 20$\%$ and 23$\%$ on the precision measurement of $\sin^{2}\theta_{23}$ and $|\Delta m^{2}_{32}|$ parameters respectively at 1$\sigma$ level over muon energy and muon direction analysis \cite{trk}. These results shows that the inclusion of hadron information together with  muon information significantly improves the capability of ICAL detector for the estimation of oscillation parameters. These results may further be improved by including the neutrino direction in the $\chi^{2}$ definition, a work under progress in INO collaboration.   

\section{Conclusions}
The magnetised ICAL detector at INO has a potential to reveal several neutrino properties, especially the mass hierarchy of the neutrino through earth matter effect and a comprehensive information on neutrino oscillation parameters. We have studied the ICAL detector capability for the precise measurement of atmospheric neutrino oscillation parameters using neutrino energy and muon angle as observables. A Monte Carlo simulation using NUANCE generated neutrino data for 10 years exposure of ICAL detector has been carried out. Analysis has been performed in the framework of three neutrino flavor mixing and by taking earth matter effect into account. A marginalised $\chi^{2}$ analysis in fine bins of reconstructed neutrino energy and muon angle has been performed. Realistic detector resolutions and efficiencies, generated from ICAL detector simulation have been utilised. The effect of various systematic uncertainties have also been included in the analysis. We conclude that by using reconstructed neutrino energy and muon direction there is an average improvement of about 20$\%$ on the precision measurement of both the parameters ($\sin^{2}\theta_{23}$ and $|\Delta m^{2}_{32}|$) over muon energy, muon angle analysis \cite{trk}. Moreover, this study is also a demonstration of the fact that the ICAL experiment has the capability of harnessing hadron information to further improve the measurement of oscillation parameters.

\section{Acknowledgments}
We thank all the INO collaborators especially physics analysis group members for  important discussions. We thank Anushree Ghosh, Tarak Thakore for their continuous help during the analysis. We are also grateful to  N. Mondal, Amol Dighe, D. Indumathi and S. Choubey for their important comments and suggestions throughout this work. Thanks to INO simulation group for providing the ICAL detector response for muons and hadrons. We also thank Department of Science and Technology (DST), Council of Scientific and Industrial Research (CSIR) and University of Delhi R$\&$D grants for providing the financial support for this research.

\clearpage

\end{document}